\documentclass[11pt]{amsart}
\usepackage{amssymb,amsthm,amsmath,epsfig,latexsym}
\usepackage{calc,times,verbatim}
\usepackage{geometry}
\usepackage{graphicx}
\usepackage{xytree}
\usepackage[T1]{fontenc}
\usepackage{extarrows}
\usepackage{color}
\usepackage{bm}
\usepackage{mathtools}
\begin{document}

\title{Finding the minimum distance and decoding linear codes with the Gaussian elimination method}
\author{Tianshuo Yang}

\renewcommand{\thefootnote}{}\footnote{Tianshuo Yang: Department of Robotics Engineering, Widener University, Chester, PA 19013, USA, Email: tyang3@widener.edu.}

\begin{abstract}
We propose an algorithm using the Gaussian elimination method to find the minimal Hamming distance and decode received messages of linear codes. This algorithm is easy to implement as it requires no Gr\"{o}bner bases to compute solutions for systems of polynomial equations. 
\end{abstract}
\maketitle

\vspace{-0.2in}

\section{Introduction}

Let $\mathbb{K}$ be a field and $\mathbb{K}^n$ the $n$-dimensional $\mathbb{K}$-vector space. A {\it linear code} $C$ of length $n$ over $\mathbb{K}$ is a subspace of $\mathbb{K}^n$. Let $\bm{v}=\left[\begin{array}{ccc}v_1&\ldots&v_n\end{array}\right]$ and $\bm{w}=\left[\begin{array}{ccc}w_1&\ldots&w_n\end{array}\right]$ be two vectors in $\mathbb{K}^n$. Recall the {\it (Hamming) distance} from $\bm{v}$ to $\bm{w}$, denoted by $d(\bm{v}, \bm{w})$, is defined to be the number of positions at which $\bm{v}$ and $\bm{w}$ differ. The {\it (Hamming) weight} of $\bm{v}$, denoted by ${\rm wt}(\bm{v})$, is defined as $d(\bm{v}, \bm{0})$, i.e., the number of entries in $\bm{v}$ that are different from zero. Let $C$ be a code containing at least two codewords. The {\it minimum (Hamming) distance} of $C$ is defined as $d(C)={\rm min}\{d(\bm{v}, \bm{w})\,|\, \bm{v}, \bm{w}\in C, \bm{v}\neq \bm{w}\}$ and the {\it minimum (Hamming) weight} of $C$ is defined as ${\rm wt}(C)={\rm min}\{{\rm wt} (\bm{v})\,|\, \bm{v}\in C, \bm{v}\neq \bm{0}\}.$ In the case that $C$ is a linear code, one has $d(C)={\rm wt}(C)$. By an $[n, k, d]$ linear code $C$ we mean a $k$-dimensional subspace of $\mathbb{K}^n$ such that the minimum distance $d(C)=d$. Under the standard coding theory, the code $C$ can detect up to $d-1$ errors and correct up to $\left\lfloor\frac{d-1}{2}\right\rfloor$ errors. Thus, determining the value of $d$ is critical for understanding the error detection/correction capability of $C$. However, Vardy \cite{V} showed that for general linear codes, computing the minimum distance is an $NP$-hard problem and the corresponding decision problem is $NP$-complete. Hence any general algorithm for computing the minimum distance will run in super polynomial time unless $P=NP$.

Historical techniques of decoding and computing the minimal distance reduce the problems to systems of polynomial equations in several variables over finite fields and then use Gr\"{o}bner bases to solve them (see for example \cite{BP}, \cite{DP1}, and \cite{DP2}).
Gr\"{o}bner bases can be computed via computer algebra packages such as Axiom, CoCoA, Gap, Macaulay, Singular, etc. The complexity of computing them is exponential in the case of a finite set of solutions. Motivated by the work of M. De Boer and R. Pellikaan \cite{DP1}, B. Anzis and S. Toh\u{a}neanu \cite{AT}, and R. Burity, S. Toh\u{a}neanu and Y. Xie \cite{BTX}, we reduce the problem of decoding and computing the minimal distance to systems of linear equations in several variables and use Gaussian elimination in linear algebra to compute their solutions.

This paper is structured as follows: In next section, we propose an algorithm using Gaussian elimination for computing the minimal distance and codewords with minimal distance for a linear code. Then we provide a similar algorithm to decode received messages of a linear code in Section 3. After that, we show examples to illustrate our algorithms. In the last section, we close the paper with concluding remarks.

\section{The minimal Hamming distance}

Let $C$ be an $[n, k, d]$ linear code over the field $\mathbb{K}$. Since $C$ is a $k$-dimensional subspace of the vector space $\mathbb{K}^n$, one can use a basis of $C$ to form a generating matrix of rank $k$
$$
G=\left[\begin{array}{cccc}a_{11}&a_{12}&\cdots&a_{1n}\\
a_{21}&a_{22}&\cdots&a_{2n}\\
\vdots&\vdots&&\vdots\\
a_{k1}&a_{k2}&\cdots&a_{kn}\end{array}\right],
$$
where $a_{ij}\in \mathbb{K}$. Observe $C$ is the image of the injective linear map $\phi: \mathbb{K}^k \xrightarrow{G} \mathbb{K}^n$ via $\phi(\bm{x})=\bm{x}G$ for $\bm{x}\in \mathbb{K}^k$. Assume $G$ is {\it nondegenerate}, i.e., none of the columns of $G$ is the zero column vector in $\mathbb{K}^k$.

Let $R=\mathbb{K}[x_1, \ldots, x_k]$ be a polynomial ring of $k$ variables over the field $\mathbb{K}$. Observe the $n$ columns in the generating matrix $G$ define $n$ nonzero homogeneous linear forms $\ell_j=\sum_{i=1}^ka_{ij}x_i$ $1\leq j\leq n$, in $R$.  These linear forms $\ell_1, \ldots, \ell_n$ are called the {\it defining linear forms} for the linear code $C$. For $1\leq a\leq n$, the {\it ideal generated by $a$-fold products of linear forms of $C$} is defined as
$$
I_a(C)=\langle \{\ell_{i_1}\cdots\ell_{i_a}\,|\, 1\leq i_1<\cdots<i_a\leq n\}\rangle.
$$

Let $\mathbb{P}_{\mathbb{K}}^{k-1}$ be the projective $(k-1)$-space. Recall the projective variety defined by a homogeneous ideal $J\subset R$ is $V(J)=\{P\in \mathbb{P}_{\mathbb{K}}^{k-1} \,|\, f(P)=0 \,\, \mbox{for all} \,\, f\in J\}$.
Let $\bm{v}$ be a nonzero codeword in $C$. Then $\bm{v}=\bm{x}G$, where $\bm{0}\neq\bm{x}\in \mathbb{K}^k$. Observe ${\rm wt}(\bm{v})\leq e$ if and only if all products of $e+1$ distinct entries of $\bm{v}$ are zero. This means $\bm{x}$ is a nonzero solution for all of the equations $\ell_{i_1}\cdots\ell_{i_{e+1}}=0$, where $1\leq i_1<\cdots<i_{e+1}\leq n$, i.e., $\bm{x}\in V\left(I_{e+1}(C)\right)$. One has that (see \cite{DP1})
$$
V\left(I_a(C)\right)=\{\bm{x}\in \mathbb{P}_{\mathbb{K}}^{k-1} \,|\, {\rm wt}(\bm{v})<a \,\, {\rm with} \,\, \bm{v}=\bm{x}G\},
$$
and
$$
d={\rm min}\{a \,|\, V\left(I_{a+1}(C)\right)\neq \emptyset\}.
$$

Let $\Gamma(C)$ be the set of all linear prime ideals generated by linear forms in $\{\ell_1, \ldots, \ell_n\}$. Let $\mathfrak{p}\in\Gamma(C)$ and $\nu_C(\mathfrak{p})$ be the number of linear forms in $\{\ell_1, \ldots, \ell_n\}$ that belong to $\mathfrak{p}$. The irrelevant maximal ideal $\mathfrak{m}=\langle x_1, \ldots, x_k\rangle=\langle\ell_1, \ldots, \ell_n\rangle$ and $\nu_C(\mathfrak{m})=n$. By \cite{BTX}, for $1\leq a\leq n$, the ideal $I_a(C)$ has the primary decomposition
$$
I_a(C)=\cap_{\mathfrak{p}\in\Gamma(C)}\mathfrak{p}^{a-n+\nu_C(\mathfrak{p})},
$$
where if $a-n+\nu_C(p)\leq 0$, then the corresponding component is replaced with the ring $R$.

Consider the case $a=1$ and $I_1(C)=\langle\ell_1, \ldots, \ell_n\rangle=\langle x_1, \ldots, x_k\rangle$. One has $V\left(I_{1}(C)\right)=\emptyset$ as $\ell_1, \ldots, \ell_n$ span a $k$-dimensional vector space so that the homogeneous equations $\ell_1=0, \ldots, \ell_n=0$ have only the trivial solution $\bm{0}$ in $\mathbb{K}^k$. By the above results, to compute the minimal distance $d$, one needs to find the maximum number of linear forms in $\{\ell_1, \ldots, \ell_n\}$ that span a $k-1$-dimensional vector space.
In other words, $d$ is the minimal number of linear forms we have to delete from $\{\ell_1, \ldots, \ell_n\}$ so that the remaining linear forms span a $k-1$-dimensional vector space. The question of whether a set of linear forms span a $(k-1)$-dimensional vector space can be answered by reducing the matrix formed by the coefficients of these linear forms to a row echelon matrix and looking at the rank.
Since we need at least $k-1$ linear forms to span a $k-1$-dimensional vector space, one has that $1\leq d\leq n-k+1$. We state the algorithm using Gaussian elimination for computing the minimal distance and codewords with minimal distance in the following:
\bigskip

{\bf Algorithm 2.1}
\medskip

\begin{enumerate}
\item Input a generating matrix $G$ of size $k\times n$
\item For $1\leq j\leq n-k+1$
\item \quad Compute $S_j=\left\{\{i_1, \ldots, i_j\} \,|\, 1\leq i_1<i_2<\cdots<i_j\leq n\right\}$
\item \quad For each $\{i_1, \ldots, i_j\}\in S_j$
\item \quad\quad Let $G_{i_1\cdots i_j}$ be the matrix obtained from $G$ by deleting $j$ columns $i_1, \ldots, i_j$
\item \quad\quad Reduce $G_{i_1\cdots i_j}^T$ to a row echelon matrix $H$. Compute ${\rm rank}(H)$.
\item \quad\quad If ${\rm rank}(H)<k$
\item[] \quad\quad\quad Solve the system of linear equations $H\bm{x}=\bm{0}$ in the projective space $\mathbb{P}_{\mathbb{K}}^{k-1}$
\item[] \quad\quad\quad Store the solutions in the set $X$
\item \quad If $X\neq \emptyset$, let $d=j$ and $Y=X^TG$. Return $d$, $X$, and $Y$.
\end{enumerate}
\bigskip

In the above algorithm, observe $|S_j|=\binom{n}{j}$ for each $1\leq j\leq n-k+1$. Thus the complexity of this algorithm is large as the number of all possible subsets of $\{1, \ldots, n\}$ grows exponentially. However, the algorithm is easy to implement and requires no symbolic computations in computer algebra systems.

\section{Decoding linear codes}

Let $C$ be an $[n, k, d]$ linear code with a generating matrix $G$ as in Section 2. Suppose that a codeword $\bm{w}=\left[\begin{array}{ccc}w_1&\ldots&w_n\end{array}\right]\in \mathbb{K}^n$ is received. The most commonly used rule for decoding $\bm{w}$ is to find the codeword $\bm{v}\in C$ which minimizes ${\rm wt}(\bm{w}-\bm{v})$ (i.e., $\bm{v}$ is the nearest neighbor of $\bm{w}$ in $C$), and decode $\bm{w}$ to $\bm{v}$. Of course, a codeword $\bm{w}\not\in C$ might have more than one nearest neighbors. In this case the nearest neighbor decoding rule fails. As we have mentioned above, the minimal distance $d$ determines the error detection/correction capability of $C$ as it can detect up to $d-1$ errors and correct up to $\left\lfloor\frac{d-1}{2}\right\rfloor$ errors.

Traditionally one translates the syndrome decoding algorithm into the language of varieties (called syndrome varieties) and use computational algebraic techniques (such as Gr\"{o}bner bases) to find the error and the nearest neighbor of a received word (see for example \cite{BP}, \cite{DP1}, and \cite{DP2}). In 2015, it was shown in \cite{AT} that any error with weight up to $\left\lfloor\frac{d-1}{2}\right\rfloor$ in data transmission can be computed as the codeword of minimum weight of a new linear code $C\left(\bm{w}\right)$ with a generating matrix
$$
G\left(\bm{w}\right)=\left[\begin{array}{cccc}a_{11}&a_{12}&\cdots&a_{1n}\\
\vdots&\vdots&&\vdots\\a_{k1}&a_{k2}&\cdots&a_{kn}\\
w_{1}&w_{2}&\cdots&w_{n}\end{array}\right],
$$
which is created from the generator matrix $G$ of $C$ by augmenting the received word $\bm{w}$ as a new row (a code with such a generating matrix is called an {\it augmented code}). Let $d\left(\bm{w}\right)={\rm min}\{{\rm wt}(\bm{e})\,|\,\bm{e}\in \mathbb{K}^n \,\, {\rm with} \,\, \bm{w}-\bm{e}\in C\}={\rm min}\{{\rm wt}(\bm{e})\,|\,\bm{e}\in C\left(\bm{w}\right)\}$. Then one can compute the codeword of minimum weight by solving the ideal $I_{d\left(\bm{w}\right)+1}\left(C\left(\bm{w}\right)\right)$ using Gr\"{o}bner bases or by finding a primary decomposition of this ideal. Both methods require symbolic computations.

Applying similar idea to the new linear code $C\left(\bm{w}\right)$ as in Section 2, we can use Gaussian elimination to compute the codeword of minimal weight in $C\left(\bm{w}\right)$. Hence we obtain the following algorithm for decoding a linear code $C$ with minimal distance $d$:
\bigskip

{\bf Algorithm 3.1}
\medskip

\begin{enumerate}
\item Input a generating matrix $G$ of size ${k\times n}$ and a received message $\bm{w}$
\item Let $G\left(\bm{w}\right)$ be the matrix obtained by appending $\bm{w}$ to $G$ in the last row
\item For $1\leq j\leq \left\lfloor\frac{d-1}{2}\right\rfloor$
\item \quad Compute $S_j=\left\{\{i_1, \ldots, i_j\} \,|\, 1\leq i_1<i_2<\cdots<i_j\leq n\right\}$
\item \quad For each $\{i_1, \ldots, i_j\}\in S_j$
\item \quad\quad Let $G\left(\bm{w}\right)_{i_1\cdots i_j}$ be the matrix obtained from $G\left(\bm{w}\right)$ by deleting $j$ columns $i_1, \ldots, i_j$
\item \quad\quad Reduce $G\left(\bm{w}\right)_{i_1\cdots i_j}^T$ to a row echelon matrix $H$. Compute ${\rm rank}(H)$.
\item \quad\quad If ${\rm rank}(H)<k+1$
\item[] \quad\quad\quad Solve the system of equations $H\bm{x}=\bm{0}$ in the projective space $\mathbb{P}_{\mathbb{K}}^{k}$
\item[] \quad\quad\quad Let $\bm{x}$ be the nonzero solution (with the last entry $=1$)
\item[] \quad\quad\quad Let $\bm{e}=\bm{x}^TG\left(\bm{w}\right)$ and $\bm{v}=\bm{w}-\bm{e}$
\item[] \quad\quad\quad Return $\bm{e}$ and $\bm{v}$
\item Return $\bm{w}$ is non-decodable
\end{enumerate}

\section{Examples}

In this section, we provide examples to illustrate the above algorithms for computing the minimal distance and decoding linear codes.
\medskip

{\bf Example 4.1} \, Let $\mathbb{K}=\mathbb{F}_2$. Consider the linear code $C$ with a generating matrix
$$
G=\left[\begin{array}{cccccc}1&0&0&1&1&0\\
0&1&0&1&0&1\\0&0&1&0&1&1\end{array}\right].
$$
This code has 6 homogeneous linear forms $x_1, x_2, x_3, x_1+x_2, x_1+x_3, x_2+x_3$ generated by the columns of $G$. Applying Algorithm 2.1, we found that $d=d(C)=3$ and there are 4 linear prime ideals of height 2; each is generated by $n-d=6-3=3$ linear forms: $\mathfrak{p}_1=\langle x_1+x_2, x_1+x_3, x_2+x_3\rangle$, $\mathfrak{p}_2=\langle x_2, x_3, x_2+x_3\rangle$, $\mathfrak{p}_3=\langle x_1, x_3, x_1+x_3\rangle$, and $\mathfrak{p}_4=\langle x_1, x_2, x_1+x_2\rangle$. To find $\cup_{i=1}^4V\left(\mathfrak{p}_i\right)$, we solve the 4 homogeneous linear systems with augmented matrices (formed by the coefficients of linear forms in $\mathfrak{p}_i, 1\leq i\leq 4$)
$$
\left[\begin{array}{ccc|c}1&1&0&0\\1&0&1&0\\0&1&1&0\end{array}\right],
\left[\begin{array}{ccc|c}0&1&0&0\\0&0&1&0\\0&1&1&0\end{array}\right],
\left[\begin{array}{ccc|c}1&0&0&0\\0&0&1&0\\1&0&1&0\end{array}\right],
\left[\begin{array}{ccc|c}1&0&0&0\\0&1&0&0\\1&1&0&0\end{array}\right],
$$
and obtain 4 nonzero solutions $\bm{x}_1=\left[\begin{array}{ccc}1&1&1\end{array}\right]^T$, $\bm{x}_2=\left[\begin{array}{ccc}1&0&0\end{array}\right]^T$,
$\bm{x}_3=\left[\begin{array}{ccc}0&1&0\end{array}\right]^T$, and $\bm{x}_4=\left[\begin{array}{ccc}0&0&1\end{array}\right]^T$. This yields 4 codewords in $C$ of minimal weight $3$:
\begin{eqnarray*}
\bm{y}_1&=&\bm{x}_1^TG=\left[\begin{array}{cccccc}1&1&1&0&0&0\end{array}\right],\\ \bm{y}_2&=&\bm{x}_2^TG=\left[\begin{array}{cccccc}1&0&0&1&1&0\end{array}\right],\\ \bm{y}_3&=&\bm{x}_3^TG=\left[\begin{array}{cccccc}0&1&0&1&0&1\end{array}\right],\\
\bm{y}_4&=&\bm{x}_4^TG=\left[\begin{array}{cccccc}0&0&1&0&1&1\end{array}\right].
\end{eqnarray*}
Hence the outputs of Algorithm 2.1 are
$$d=3,$$
$$X=\left\{\left[\begin{array}{ccc}1&1&1\end{array}\right]^T, \left[\begin{array}{ccc}1&0&0\end{array}\right]^T, \left[\begin{array}{ccc}0&1&0\end{array}\right]^T, \left[\begin{array}{ccc}0&0&1\end{array}\right]^T\right\},$$
and
$$
Y=\left\{\begin{array}{cc}\left[\begin{array}{cccccc}1&1&1&0&0&0\end{array}\right]&
\left[\begin{array}{cccccc}1&0&0&1&1&0\end{array}\right]\\
&\\
\left[\begin{array}{cccccc}0&1&0&1&0&1\end{array}\right]&\left[\begin{array}{cccccc}0&0&1&0&1&1\end{array}\right]
\end{array}\right\}.
$$

\noindent
Indeed, there are 8 codewords in $C$, i.e.,
$$
C=\left\{\begin{array}{cccc}
\left[\begin{array}{cccccc}0&0&0&0&0&0\end{array}\right]&
\left[\begin{array}{cccccc}1&0&0&1&1&0\end{array}\right]&
\left[\begin{array}{cccccc}0&1&0&1&0&1\end{array}\right]&
\left[\begin{array}{cccccc}0&0&1&0&1&1\end{array}\right]\\
&\\
\left[\begin{array}{cccccc}1&1&0&0&1&1\end{array}\right]&
\left[\begin{array}{cccccc}1&0&1&1&0&1\end{array}\right]&
\left[\begin{array}{cccccc}0&1&1&1&1&0\end{array}\right]&
\left[\begin{array}{cccccc}1&1&1&0&0&0\end{array}\right]
\end{array}\right\}.
$$
One can see that $d(C)=3$ and there are 4 codewords in $C$ of minimal weight $3$.

Since $d(C)=3$, the linear code $C$ can fix one error. Suppose the codeword $\bm{w}=\left[\begin{array}{cccccc}0&1&1&1&0&0\end{array}\right]$ is received. Then we have the augumented code $C(\bm{w})$ with the augumented matrix
$$
G(\bm{w})=\left[\begin{array}{cccccc}1&0&0&1&1&0\\
0&1&0&1&0&1\\0&0&1&0&1&1\\0&1&1&1&0&0\end{array}\right].
$$
The augumented code $C(\bm{w})$ has 6 homogeneous linear forms $x_1, x_2+x_4, x_3+x_4, x_1+x_2+x_4, x_1+x_3, x_2+x_3$ generated by the columns of $G(\bm{w})$. Applying Algorithm 3.1, we have $d\left(C(\bm{w})\right)=1$ and there is a unique linear prime ideal of height $3$ generated by $n-d=6-1=5$ linear forms: $\mathfrak{p}=\langle x_1, x_2+x_4, x_3+x_4, x_1+x_2+x_4, x_2+x_3\rangle$. 
Again to find $V\left(\mathfrak{p}\right)$, we solve the homogeneous linear system with augmented matrix (formed by the coefficients of linear forms in $\mathfrak{p}$)
$$
\left[\begin{array}{cccc|c}1&0&0&0&0\\0&1&0&1&0\\0&0&1&1&0\\1&1&0&1&0\\0&1&1&0&0\end{array}\right]
$$
and obtain the unique nonzero solution $\bm{x}=\left[\begin{array}{cccc}0&1&1&1\end{array}\right]^T$. The error codeword $\bm{e}=\bm{x}^TG(\bm{w})=\left[\begin{array}{cccccc}0&0&0&0&1&0\end{array}\right]$. Hence the codeword in $C$ with minimal distance to $\bm{w}$ is
$$\bm{v}=\bm{w}-\bm{e}=\left[\begin{array}{cccccc}0&1&1&1&1&0\end{array}\right].$$
The outputs of Algoritm 3,1 are $\bm{e}=\left[\begin{array}{cccccc}0&0&0&0&1&0\end{array}\right]$ and $\bm{v}=\left[\begin{array}{cccccc}0&1&1&1&1&0\end{array}\right]$, and we decode $\bm{w}$ to $\bm{v}$.
\medskip

{\bf Example 4.2} \, Let $\mathbb{K}=\mathbb{F}_2$. Consider the $[7, 4]$ cyclic code $C$ generated by the polynomial $g(x)=1+x^2+x^3$. Then this code has a generating matrix (note that we identify a vector with a polynomial)
$$
G=\left[\begin{array}{c}g(x)\\xg(x)\\x^2g(x)\\x^{3}g(x)\end{array}\right]=
\left[\begin{array}{ccccccc}1&0&1&1&0&0&0\\
0&1&0&1&1&0&0\\0&0&1&0&1&1&0\\0&0&0&1&0&1&1
\end{array}\right].
$$
This code has 7 homogeneous linear forms $x_1, x_2, x_1+x_3, x_1+x_2+x_4, x_2+x_3, x_3+x_4, x_4$ generated by the columns of $G$. Applying Algorithm 2.1, we found that $d=d(C)=3$ and there are 7 linear prime ideals of height 3 generated by $n-d=7-3=4$ linear forms: $\mathfrak{p}_1=\langle x_1, x_2, x_1+x_3, x_2+x_3\rangle$, $\mathfrak{p}_2=\langle x_1, x_2, x_1+x_2+x_4, x_4\rangle$, $\mathfrak{p}_3=\langle x_1, x_1+x_3, x_3+x_4, x_4\rangle$,
$\mathfrak{p}_4=\langle x_1, x_1+x_2+x_4, x_2+x_3, x_3+x_4\rangle$, $\mathfrak{p}_5=\langle x_2, x_1+x_3, x_1+x_2+x_4, x_3+x_4\rangle$, $\mathfrak{p}_6=\langle x_2, x_2+x_3, x_3+x_4, x_4\rangle$, and $\mathfrak{p}_7=\langle x_1+x_3, x_1+x_2+x_4, x_2+x_3, x_4\rangle$.

To find $\cup_{i=1}^7V\left(\mathfrak{p}_i\right)$, we solve the 7 homogeneous linear systems with augmented matrices (formed by the coefficients of linear forms in $\mathfrak{p}_i, 1\leq i\leq 7$)
$$
\left[\begin{array}{cccc|c}1&0&0&0&0\\0&1&0&0&0\\1&0&1&0&0\\0&1&1&0&0\end{array}\right],
\left[\begin{array}{cccc|c}1&0&0&0&0\\0&1&0&0&0\\1&1&0&1&0\\0&0&0&1&0\end{array}\right],
\left[\begin{array}{cccc|c}1&0&0&0&0\\1&0&1&0&0\\0&0&1&1&0\\0&0&0&1&0\end{array}\right],
\left[\begin{array}{cccc|c}1&0&0&0&0\\1&1&0&1&0\\0&1&1&0&0\\0&0&1&1&0\end{array}\right],
$$
$$
\left[\begin{array}{cccc|c}0&1&0&0&0\\1&0&1&0&0\\1&1&0&1&0\\0&0&1&1&0\end{array}\right],
\left[\begin{array}{cccc|c}0&1&0&0&0\\0&1&1&0&0\\0&0&1&1&0\\0&0&0&1&0\end{array}\right],
\left[\begin{array}{cccc|c}1&0&1&0&0\\1&1&0&1&0\\0&1&1&0&0\\0&0&0&1&0\end{array}\right],
$$
and obtain 7 nonzero solutions, i.e., $X=\{\bm{x}_i \,|\, 1\leq i\leq 7\}$, where 
$\bm{x}_1=\left[\begin{array}{cccc}0&0&0&1\end{array}\right]^T$, 
$\bm{x}_2=\left[\begin{array}{cccc}0&0&1&0\end{array}\right]^T$,
$\bm{x}_3=\left[\begin{array}{cccc}0&1&0&0\end{array}\right]^T$,
$\bm{x}_4=\left[\begin{array}{cccc}0&1&1&1\end{array}\right]^T$,
$\bm{x}_5=\left[\begin{array}{cccc}1&0&1&1\end{array}\right]^T$,
$\bm{x}_6=\left[\begin{array}{cccc}1&0&0&0\end{array}\right]^T$,
and $\bm{x}_7=\left[\begin{array}{cccc}1&1&1&0\end{array}\right]^T$. 
This yields 7 codewords in $C$ of minimal weight $3$, i.e.,
$Y=\{\bm{y}_i \,|\, 1\leq i\leq 7\}$, where
\begin{eqnarray*}
\bm{y}_1&=&\bm{x}_1^TG=\left[\begin{array}{ccccccc}0&0&0&1&0&1&1\end{array}\right],\\ \bm{y}_2&=&\bm{x}_2^TG=\left[\begin{array}{ccccccc}0&0&1&0&1&1&0\end{array}\right],\\ \bm{y}_3&=&\bm{x}_3^TG=\left[\begin{array}{ccccccc}0&1&0&1&1&0&0\end{array}\right],\\
\bm{y}_4&=&\bm{x}_4^TG=\left[\begin{array}{ccccccc}0&1&1&0&0&0&1\end{array}\right],\\
\bm{y}_5&=&\bm{x}_4^TG=\left[\begin{array}{ccccccc}1&0&0&0&1&0&1\end{array}\right],\\
\bm{y}_6&=&\bm{x}_4^TG=\left[\begin{array}{ccccccc}1&0&1&1&0&0&0\end{array}\right],\\
\bm{y}_7&=&\bm{x}_4^TG=\left[\begin{array}{ccccccc}1&1&0&0&0&1&0\end{array}\right].
\end{eqnarray*}
By computing the 16 codewords in $C$, one can verify that $d(C)=3$ and there are 7 codewords of minimal weight $3$.

Again since $d(C)=3$, the cyclic code $C$ can only fix one error. Suppose $\bm{w}=\left[\begin{array}{ccccccc}1&1&0&1&0&1&1\end{array}\right]$ is the received codeword. Then we have the augumented code $C(\bm{w})$ with the augumented matrix
$$
G(\bm{w})=\left[\begin{array}{ccccccc}1&0&1&1&0&0&0\\
0&1&0&1&1&0&0\\0&0&1&0&1&1&0\\0&0&0&1&0&1&1\\1&1&0&1&0&1&1
\end{array}\right].
$$
This augumented code $C(\bm{w})$ has 7 homogeneous linear forms $x_1+x_5, x_2+x_5, x_1+x_3, x_1+x_2+x_4+x_5, x_2+x_3, x_3+x_4+x_5, x_4+x_5$ generated by the columns of $G(\bm{w})$. Applying Algorithm 3.1, we have $d\left(C(\bm{w})\right)=1$ and there is a unique linear prime ideal of height $4$ generated by $n-d=7-1=6$ linear forms: $\mathfrak{p}=\langle x_1+x_5, x_2+x_5, x_1+x_3, x_1+x_2+x_4+x_5, x_2+x_3, x_4+x_5\rangle$. 
To find $V\left(\mathfrak{p}\right)$, we solve the homogeneous linear system with augmented matrix (formed by the coefficients of linear forms in $\mathfrak{p}$)
$$
\left[\begin{array}{ccccc|c}1&0&0&0&1&0\\0&1&0&0&1&0\\1&0&1&0&0&0\\1&1&0&1&1&0\\0&1&1&0&0&0\\0&0&0&1&1&0\end{array}\right]
$$
and obtain the unique nonzero solution $\bm{x}=\left[\begin{array}{ccccc}1&1&1&1&1\end{array}\right]^T$. The error codeword $\bm{e}=\bm{x}^TG(\bm{w})=\left[\begin{array}{ccccccc}0&0&0&0&0&1&0\end{array}\right]$. Hence the codeword in $C$ with minimal distance to $\bm{w}$ is
$$\bm{v}=\bm{w}-\bm{e}=\left[\begin{array}{ccccccc}1&1&0&1&0&0&1\end{array}\right].$$
The outputs of Algoritm 3,1 are $\bm{e}=\left[\begin{array}{ccccccc}0&0&0&0&0&1&0\end{array}\right]$ and $\bm{v}=\left[\begin{array}{ccccccc}1&1&0&1&0&0&1\end{array}\right]$, and we decode $\bm{w}$ to $\bm{v}$.
\bigskip

\section{Concluding remarks}

The purpose of this paper is to propose a practical algorithm for computing the minimal distance and decoding general linear codes without using symbolic computations in computer algebra systems. The computational complexity of this algorithm is large as one cannot hope for a polynomial algorithm to decode and compute the minimal distance for general linear codes. However since the implementation is not hard, one can use it to compute certain examples to verify your intuition. This can help design new linear codes.

One can modify the above algorithm to compute other things for an $[n, k]$ linear code $C$ such as the primary decomposition of $I_a(C)$ for $1\leq i\leq n$, the weight distribution $\{\left(i, \alpha_i\right)\,|\,i=0, 1, \ldots, n\}$, where $\alpha_i$ denotes the number of codewords in $C$ of weight $i$, the polynomial $W_C(X, Y)=\sum_{i=0}^n\alpha_iX^{n-i}Y^i$, and the MacWilliams identity $W_{C^{\perp}}\left(X, Y\right)=q^{-k}W_C\left(X+(q-1)Y, X-Y\right)$, where $\mathbb{K}=\mathbb{F}_q$ is a finite field of $q$ elements and $C^{\perp}$ is the dual code of $C$.

This algorithm may be improved to reduce the computational complexity in special classes of linear codes. For example, if the generator matrix $G$ is of the form $\left[\begin{array}{ccc}G_1&\cdots&G_{s}\end{array}\right]$, where $G_1, \ldots, G_{s}$ are matrices such that the first $k$ columns of $G_j$ form the $k\times k$ identity matrix for all $j=1, \ldots, s$, we know that any linear combination of $r$ rows with non-zero coefficients gives a codeword of weight at least $rs$. Hence $s\leq d\leq n-k+1$ and we can start our loop from $j=s$. In the future, one may implement this algorithm to study special classes of linear codes such as cyclic codes, maximum distance separable (MDS) codes, BCH codes, Golay codes, etc.

\vskip 0.5in
\bibliographystyle{amsalpha}

\end{document}